\documentstyle[12pt,aaspp4,psfig]{article}

\begin{document}

\title{The Intriguing New Cataclysmic Variable KUV\,03580+0614}

\author{Paula Szkody\footnote{Based on observations
with the Apache Point Observatory (APO) 3.5m telescope, which is owned and
operated by the Astrophysical Research Consortium (ARC)}}
\affil{Department of Astronomy, University of Washington, Seattle, WA 98195\\
szkody@astro.washington.edu}

\author{Boris G\"ansicke}
\affil{Universit\"ats-Sternwarte G\"ottingen, Geismarlandstr. 11, 37073
G\"ottingen, Germany\\
boris@uni-sw.gwdg.de}

\author{Robert E. Fried}
\affil{Braeside Observatory, PO Box 906, Flagstaff AZ 86002, captain@braeside.org}

\author{Uli Heber}
\affil{Sternwarte Bamberg,
ai03@sternwarte.uni-erlangen.de}

\author{Dawn K. Erb}
\affil{Department of Astronomy, CalTech, Pasadena, CA,
dke@astro.caltech.edu}

\begin{abstract}
Photometric and spectroscopic observations of the candidate
cataclysmic variable KUV\,03580+0614, independently found from the
Hamburg survey as HS\,0357+0614, reveal this to be a cataclysmic
variable with an orbital period of 3.4 hrs. CVs with periods in this
range are usually novalikes belonging to the SW Sex or Intermediate
Polar categories. The variability, spectroscopic characteristics and
the lack of strong X-rays indicate a system similar to V795 Her, LS
Peg and AH Men.
\end{abstract}


\section{Introduction}

The Hamburg Schmidt objective prism survey (HQS; Hagen et al. 1995)
provides a magnitude-limited ($V=13-17.5$) sample of low resolution
spectra ($\sim45$\,\AA\ at $\mathrm{H}_{\gamma}$) of blue
objects. Although it was designed to pick up quasars, it has proven to
be a resource for finding blue stars (see e.g. Heber, Jordan \&
Weidemann 1991) including new cataclysmic variables (CVs).  Recent
discoveries include the UX UMa type HS\,0139+0559 (Heber, Jordan \&
Weidemann 1991), the eclipsing systems HS\,1804+6753 (EX Dra;
Billington et al. 1996) and HS\,0907+1902 (G\"ansicke et al. 2000) and
the magnetic systems HS\,1023+3900 (Reimers, Hagen \& Hopp 1999) and
HS\,0922+1333 (Reimers \& Hagen 2000).  We have independently
identified the 15.5 mag CV candidate KUV\,03580+0614 from the Kiso
Schmidt camera survey (Wegner \& Boley 1993)  as HS\,0357+0614
during the course of follow-up studies of hot star candidates from the
HQS.  Our preliminary low resolution spectrum showed prominent He\,II
$\lambda$ 4686 emission, as is also evident in the Wegner \& Boley
spectrum. This line is a characteristic signature of either a magnetic
CV (AM Her or Intermediate Polar type) or an old nova (the various
types of CV systems and their characteristics are reviewed in Warner
1995). In order to determine the specific nature of KUV\,03580+0614,
we obtained photometric and time-resolved spectroscopic observations
which provided the orbital period, the range of variability on short
and long timescales and the location of the line emission regions.

\section{Observations}

Our identification spectrum of HS\,0357+0614 (=KUV\,03580+0614),
obtained in October 1990 at the Calar Alto 3.5m telescope with the
focal reductor spectrograph with a resolution of about 7\AA\ (Figure
1), is characterized by Balmer and He\,I and He\,II emission as well
as the C,N blend at 4650\AA. The He\,II line is similar in strength to
H$\beta$, indicating the presence of a radiation field rich in
high-energy photons. The overall appearance is similar to the lower
resolution spectrum in Wegner \& Boley (1993).

Twenty higher resolution spectra over 3.5 hrs were obtained at the
Apache Point Observatory (APO) with the 3.5m telescope and Double
Imaging Spectrograph (DIS) on the night of 2000 January 11 UT. The DIS
coverage is 4200-5050\AA\ in the blue and 5800-6800\AA\ in the red
with $\sim$2.5\AA\ resolution. A 1.5 arcsec slit was used with seeing
conditions comparable to the slit width.  The spectra were analyzed in
a standard manner using IRAF.\footnote{{IRAF (Image Reduction and
Analysis Facility) is distributed by the National Optical Astronomy
Observatories, which are operated by AURA, Inc., under cooperative
agreement with the national Science Foundation.}} The images were
corrected with bias and flat fields, the spectra were extracted to one
dimension with sky subtraction, the wavelengths were calibrated using
helium, neon and argon lamps, and the fluxes were calibrated from
standard stars in the Kitt Peak standards atlas. The combined mean
blue and red spectra are shown in Figure 1. The flux differences
between the Calar Alto and APO spectra are likely due to the intrinsic
variability of the system as described in Section 4.

Differential photometry in a V filter was obtained on 3 nights at
Manastash Ridge Observatory (MRO) in 1999 Sept. and October using a
TEK CCD on the 0.76m telescope.  The bulk of our differential
photometry of  KUV\,03580+0614 was obtained in December 1999 and
January 2000 at the Braeside Observatory (BO) in Arizona, using a 0.4m
reflector equipped with a SITe\,512 CCD camera.  $B$, $V$, and $R$
magnitudes of KUV\,03580+0614 were derived relative to the comparison
stars C, C1, and C2 (Figure 2), all of which are included in the guide
star catalog and in the USNO catalog.  The last night was simultaneous
with the APO spectra.  The observations are summarized in Table 1.

\section{Photometry}

The $V$ light curves from MRO (Figure 3) and the $R$ light curves from
BO (Figure 4) show a modulation of $\sim3.5$\,h, superimposed by
flickering on time scales of $15-30$\,min. The quasi-simultaneous $B$,
$V$, and $I$ light curves (Figure 5) show that the modulation is only
marginally color-dependent. From Figure 4, it is apparent that the
amplitude of the modulation is quite variable, ranging from 0.2 mag
(Dec. 24) to 0.4 mag (Dec. 17). The bottom plot (Jan. 11) shows that
large changes are even possible from one orbit to the next.

To quantify this variation, we computed discrete fourier transforms
(DFTs) from various subsets of our photometry. These consistently
revealed a strong signal at $f\sim7\mathrm{d^{-1}}$.  Figure 6 shows
the power spectrum obtained from the combination of all our data. The
strongest signal is again found near $f=7\mathrm{d^{-1}}$ and shows a
substructure of four possible periods. The uncertainty in cycle counts
between our observation intervals prevents us from formally rejecting
any of these possible periods. We folded all our photometric data with
the four likely periods, and favor $P\mathrm{_{orb}}$=205.92\,min as
it produces the smoothest orbital light curve, even though
$P\mathrm{_{orb}}$=205.57\,min is slightly stronger in the power
spectrum. A sine fit to all the data gave a period of
$205.80\pm0.02$\,min, which is very close to our favored period.

In order to analyze the short-term fluctuations, we fit a sine with a
fixed period, $P=P\mathrm{_{orb}}$=205.92\,min, to the data of each
individual night, and pre-whitened the data with the best-fit sine
curves. We then recomputed the fourier transform on the detrended
data.  
A number of periods are present in the
range $30-100\mathrm{d}^{-1}$ which have amplitudes consistent with
the observed flickering.  This intermittent presence of QPOs in
KUV\,03580+0614 is very similar to the short-term variability of the
novalike variable TT\,Ari (Tremko et al. 1996). In TT\,Ari, which has
been considered for a long time as an intermediate polar candidate,
these QPOs where interpreted either as the Kepler period at the inner
edge of a magnetically truncated disc or as the beat between the
magnetospheric period and the white dwarf spin period (e.g. Hollander
\& van Paradijs 1992).

\section{Spectroscopy}

Throughout the 3.5 hrs of APO spectra, the emission lines are broad,
with the Balmer lines showing double and single peak components while
the He\,II line remains single-peaked.  Some representative spectra
are shown in Figure 7. We used centroid and Gaussian fitting routines
within IRAF to measure the line velocities (Table 2).  These
velocities were then fit to a sinusoid to determine the period,
semi-amplitude, and gamma velocities. The best fit with the minimum
standard deviation occurred for the He\,II line and the resulting
period (204 min) was consistent with that found from the longer
photometric dataset (above). We then fixed the period at 206 min (as
determined by the photometry) and found the best values for K and
$\gamma$ by redoing the sine fits (Table 3). The final fit and the
data for He\,II is shown in Figure 8.  After submitting our paper, we
became aware of a preprint by Thorstensen and Taylor (2001), which
used H$\alpha$ data to determine a slightly longer period of 215 min
(the cycle ambiguities in their data do not allow phasing together
their 3 nights of observation nor any of their nights with
ours). Fixing the sinusoid period at 215 min with our data produced
the same values as shown in Table 3 (with a shift of 0.04 phase).  The
period of 215 min is the second strongest peak in our power spectrum
(Figure 6). Although they are not common, there are systems known
(e.g. TT Ari) that have shorter photometric periods --~``negative
superhumps''~-- than orbital periods (Warner 1995). However, rather
than speculating further, we await more extensive spectroscopic and
photometric data to determine if the 4\% difference is real and
persistent and merely regard our phases as uncertain by 0.04.

In order to provide more information as to the nature of the accretion
in this system, we computed doppler tomograms (Horne 1991).  To
construct these, we used a fourier-filtered back-projection program
from Keith Horne that was modified for our computer and plotting needs
by Donald Hoard. Since there is no eclipse to mark the zero phasing,
we used the red-blue crossing of the emission line velocity from the
HeII\, line which would correspond to inferior conjunction of the
secondary if the emission lines originate from the white dwarf
area. For our data, this phase 0 occurs at JD 2,451,554.650.  The
resulting tomograms for H$\alpha$, H$\beta$ and He\,II are shown in
Figure 9. Each tomogram shows a completely different structure. The
He\,II shows a spot at a location close to the white dwarf (the small
offset from V$_{x}$=0 is within the accuracy of our phase
determination). The H$\alpha$ line shows a concentration at 0 velocity
with an outer ring, while the H$\beta$ emission shows a half ring,
with a lack of emission from phases 0.6-0.9. The H$\alpha$ also shows
less emission at these phases.

Since the simultaneous photometry from BO on this night (bottom plot
in Figure 4) shows a large change in amplitude of the orbital
modulation from the beginning to the end of the APO spectra, it is
possible that the changing line intensities from variable mass
transfer are causing some of the peculiarities in the tomograms. One
of the basic assumptions of tomography is that the emitting regions
are equally visible at all orbital phases so transient emission alters
the correct picture. However, since the red and blue spectra are
obtained simultaneously, this does not explain the large differences
between the lines.  The APO spectra start at phase 0.42 and end at
1.44 and the increased flux evident in the photometry and spectra
begins about phase 1.2 so phases 0.6-0.9 are in the low fluxes (and
evident as low equivalent widths in the Balmer lines during these
phases). However, phases 0.4-0.6 (where the equivalent widths are
large and there is strong emission in the tomogram (near V$_{x}$=0,
V$_{y}$=-300) is also in this low flux interval. It will require
further data obtained with less change within an orbit to sort out the
stable configuration in this system. We can conclude that the high
excitation line emission originates close to the white dwarf while the
Balmer emission is partly from a disk structure.

From the simultaneous photometry/spectra on Jan. 11, we can determine
the orbital modulation peaks at phase $\sim$0.75. This phase is near
that where a hot spot, created by the interaction of the mass transfer
stream with the disk, would typically be best viewed.  If this area is
optically thick, it may be a strong continuum source, but may not be a
source of line emission.
 
\section{Conclusions}

While the orbital period of KUV\,03580+0614 seems clearly determined
to be near 3.4 hrs, the exact nature of this novalike system is
ambiguous.  The  bulk of CVs that occupy periods between 3-4 hrs
consist of both the high mass transfer novalike SW Sex stars and the
magnetic IPs. In general, the distinguishing traits of the SW Sex
stars are strong He\,II, single-peaked emission lines and absorption
appearing in the Balmer lines at some phases (usually near 0.5).  They
are generally low X-ray emitters. The IPs usually show a consistent
spin-pulse and large hard X-ray flux, although some have QPOs and
others have low X-ray flux. They can also have strong He\,II.
KUV\,03580+0614 has not been detected in the ROSAT All Sky Survey, so
it is likely a faint X-ray source. 

The presence of a photometric modulation that may be 4\% shorter than
the orbital period, the sporadic prescence of a flickering timescale
near 20 min and large variability in the structure of the Balmer lines
throughout the orbit make this system very similar to a handful
objects in the 3-4 hr period range with unique properties such as AH
Men (Buckley et al. 1993), V795 Her (Patterson \& Skillman 1994), LS
Peg (Szkody et al. 1997) and TT Ari (Kraicheva et al. 1999).  Whatever
the affiliation of KUV\,03580+0614, its presence in the 3--4\,hrs
period range provides another chance to study the bizarre behavior of
this type of CV. These systems, characterized by rather high accretion
rates, are flanked by the period gap for $P\mathrm{_{orb}}\la3$\,hr
and by dwarf novae with much lower accretion rates, such as U\,Gem or
IP\,Peg, for $P\mathrm{_{orb}}\ga4$\,hr. Besides the empirical
determination of the accretion flow in these high $\dot{M}$ systems
between 3-4 hrs, the underlying deficiency in our understanding of CV
evolution is the cause of the high mass transfer rates - are they
related to a peculiar characteristic of the primary or to more
activity from the secondary?

\acknowledgements
We gratefully acknowledge Keith Horne and Donald Hoard for providing
the programs for Doppler tomography.

BTG was supported by DLR/BMBF grant 50\,OR\,9903\,6. The
HQS was supported by the Deutsche Forschungsgemeinschaft through
grants Re\,353/11 and Re\,353/22. Braeside Observatory acknowledges
the support of The Research Corporation, The National Science
Foundation (AST-92-180002), and the Fund of Astrophysical Research.


\begin{deluxetable}{lccccl}
\tablenum{1}
\tablewidth{0pt}
\tablecaption{Summary of Observations}
\tablehead{
\colhead{Date} & \colhead{UT Time} & \colhead{Obs} & \colhead{Data} & 
\colhead{Exp.(s)} & \colhead{Num. Obs} }
\startdata
1990 Oct 04 & 01:57 - 02:37 & CA & spectra & 900 & 1 \nl 
1999 Sep 23 & 08:52 - 12:59 & MRO & V phot & 30 & 199  \nl 
1999 Oct 16 & 11:37 - 12:24 & MRO & V phot & 30 & 83 \nl
1999 Oct 17 & 10:18 - 12:47 & MRO & V phot & 30 & 147 \nl
1999 Dec 16 & 03:04 - 10:06 & BO & R phot & 50 & 376 \nl
1999 Dec 17 & 02:31 - 10:14 & BO & R phot & 50 & 452 \nl
1999 Dec 24 & 02:22 - 09:47 & BO & R phot & 50 & 449 \nl
1999 Dec 27 & 01:23 - 09:36 & BO & B,V,I phot & 143, 75, 50 & 96, 94, 99 \nl
2000 Jan 11 & 01:54 - 08:34 & BO & R phot & 50 & 377 \nl
2000 Jan 11 & 01:35 - 05:12 & APO & spectra & 600 & 20 \nl
\enddata
\end{deluxetable}

\begin{deluxetable}{rrrr}
\tablenum{2}
\tablewidth{0pt}
\tablecaption{2000 January 11 Velocities (km/s)}
\tablehead{
\colhead{Time (UT min)} & \colhead{He $\lambda$4686} & \colhead{H$\alpha$}
& \colhead{H$\beta$} }
\startdata
97.5 & -53 & 16 & 104 \nl
105.0 & -115 & 0 & 63 \nl
114.2 & -18 & -21 & -15 \nl
125.5 & 3 & 26 & -3 \nl
136.7 & 57 & 64 & 37 \nl
148.0 & 29 & 68 & 3 \nl
159.2 & 48 & 50 & -28 \nl
170.5 & 41 & 86 & -177 \nl
181.8 & 18 & 69 & -89 \nl
193.0 & -14 & 8 & -197 \nl
204.3 & 12 & 30 & -157 \nl
215.5 & -2 & 51 & -72 \nl
226.9 & -15 & 15 & -153 \nl
238.2 & -46 & -21 & -186 \nl
249.4 & -66 & -35 & -169 \nl
260.7 & -138 & -57 & -219 \nl
271.9 & -88 & -45 & -67 \nl
283.7 & -96 & 116 & 120 \nl
295.0 & -64 & 73 & 102 \nl
306.4 & -38 & 68 & 81 \nl 
\enddata
\end{deluxetable}

\begin{deluxetable}{lcccc}
\tablenum{3}
\tablewidth{0pt}
\tablecaption{Radial Velocity Curves}
\tablehead{
\colhead{Line} & \colhead{$\gamma$} & \colhead{K} & \colhead{$\phi$} & 
\colhead{$\sigma$} }
\startdata
H$\alpha$ & 22$\pm$1 & 44$\pm$13 & -0.03$\pm$0.04 & 31 \nl
H$\beta$ & -59$\pm$2 & 117$\pm$17 & -0.27$\pm$0.03 & 50 \nl
He\,4686 & -18$\pm$0.3 & 62$\pm$6 & 0.00$\pm$0.01 & 16 \nl
\enddata
\end{deluxetable}

\begin{figure}
\figurenum {1}
\caption{The 1990 Calar Alto identification spectrum (top) and the APO 
blue (middle) and red (bottom) mean of the 20 high resolution spectra obtained
on 11 January 2000.}

\figurenum {2}
\caption{Finding chart ($8\arcmin\times8\arcmin$) for 
HS\,0357+0614 (=KUV\,03580+0614) obtained
from the Digitized Sky Survey. The coordinates of the star are
$\alpha(2000)= 04^h00^m37.1^s$ and
$\delta(2000)=+06^{\circ}22\arcmin46\arcsec$. Our three comparison
stars are labelled C, C1, and C2.}

\figurenum {3}
\caption{Sample V light curves from MRO (plotted as differential magnitudes
with respect to a comparison star) showing the 3.5 hr
modulation.}

\figurenum {4}
\caption{Sample R light curves from BO. The bottom plot is simultaneous
with the APO spectra, which end at 1554.716.}

\figurenum {5}
\caption{Quasi-simultaneous $B$, $V$, and $I$ light curves of KUV\,03580+0614.}

\figurenum {6}
\caption{Power spectrum of all the photometric data  of KUV\,03580+0614. 
The DFT of the
window function (spectral window) is shown on top of the panel.}

\figurenum {7}
\caption{A sample of the APO spectra as a function of phase showing the
structure in H$\beta$ versus the single-peaked nature of He\,II.}

\figurenum {8}
\caption{The best fit radial velocity curve to the He\,II data, with the
fit parameters of Table 3. The two open circles may be contaminated by
cosmic rays and were excluded from the fit.}

\figurenum {9}
\caption{The Doppler tomograms of a) He\,II, b) H$\alpha$ and c) H$\beta$.}
\end{figure}

\clearpage
\begin{figure}
\psfig{figure=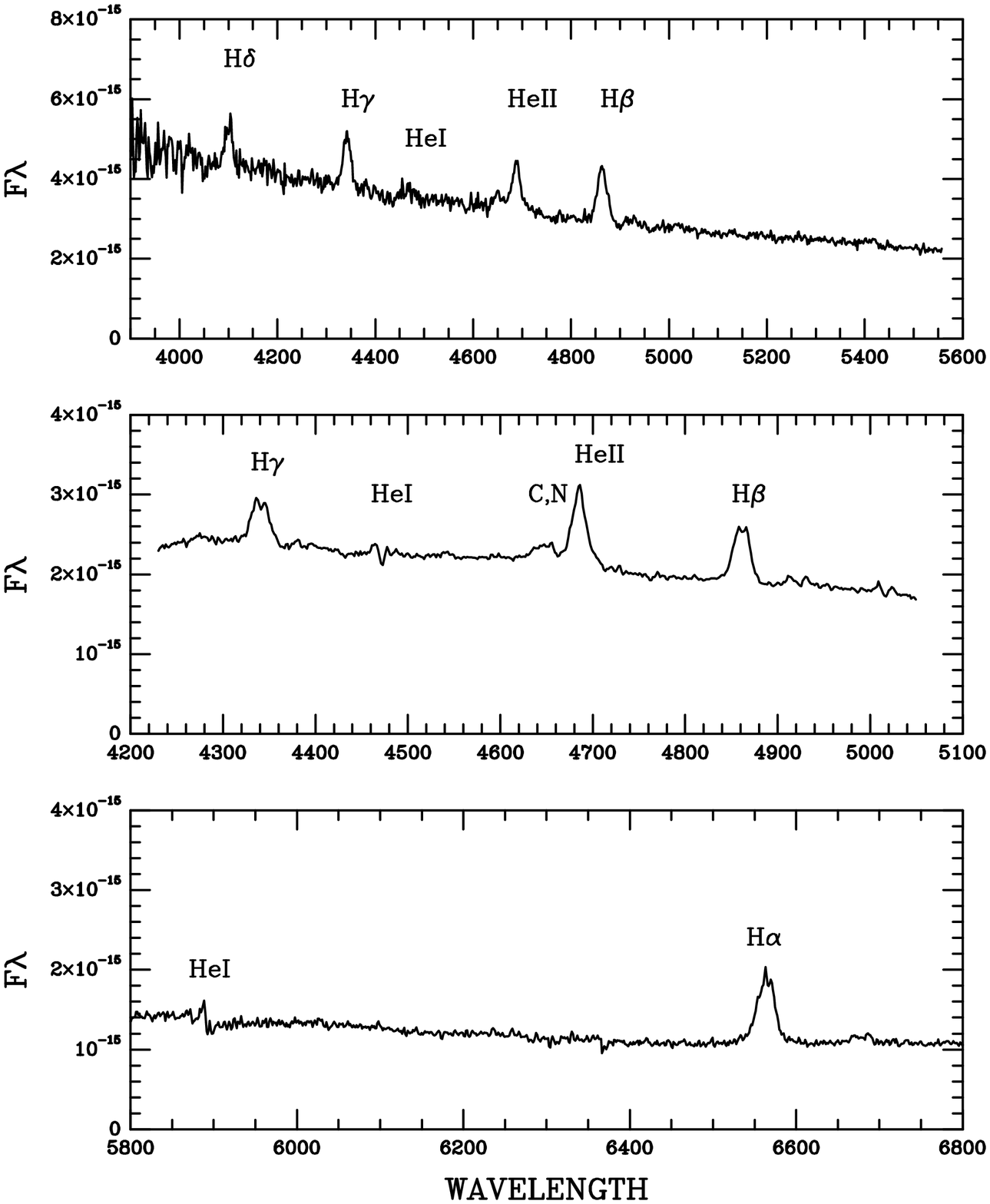,width=7in}
\end{figure}
\clearpage
\begin{figure}
\psfig{figure=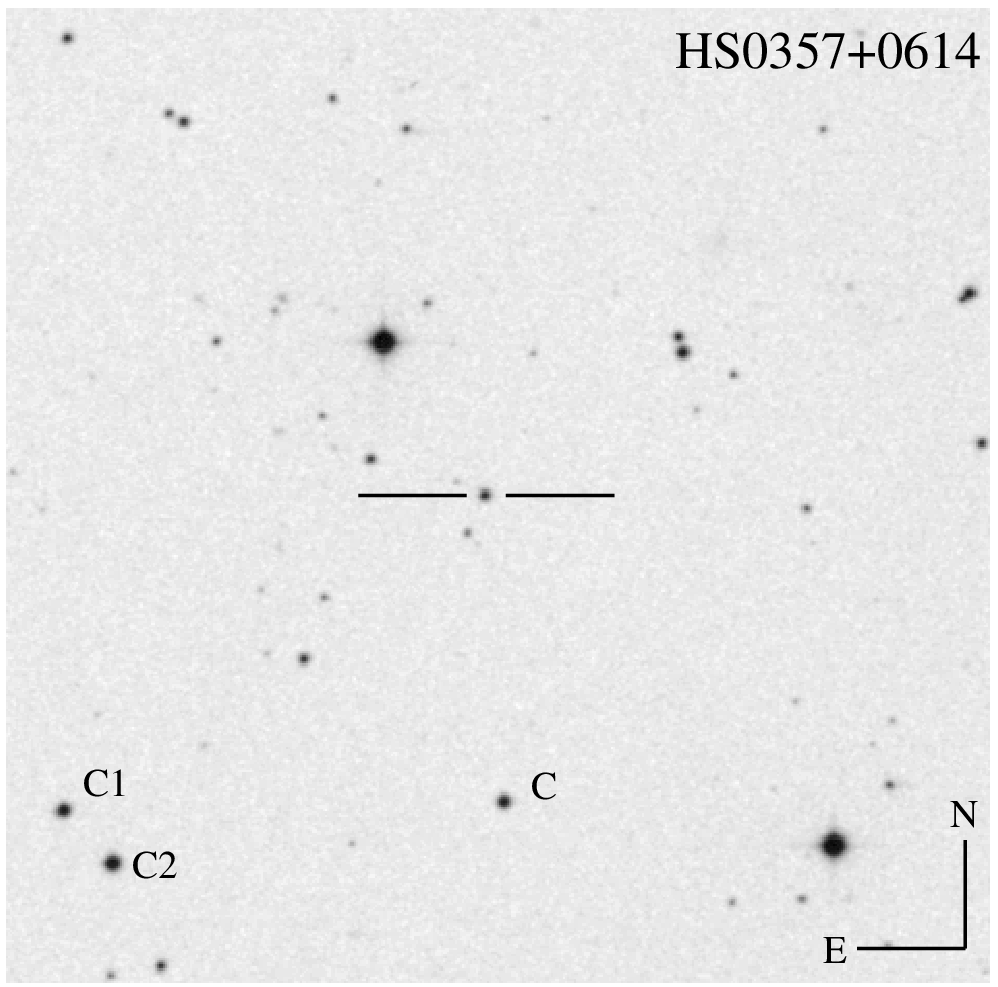,width=4in}
\end{figure}
\clearpage
\begin{figure}
\psfig{figure=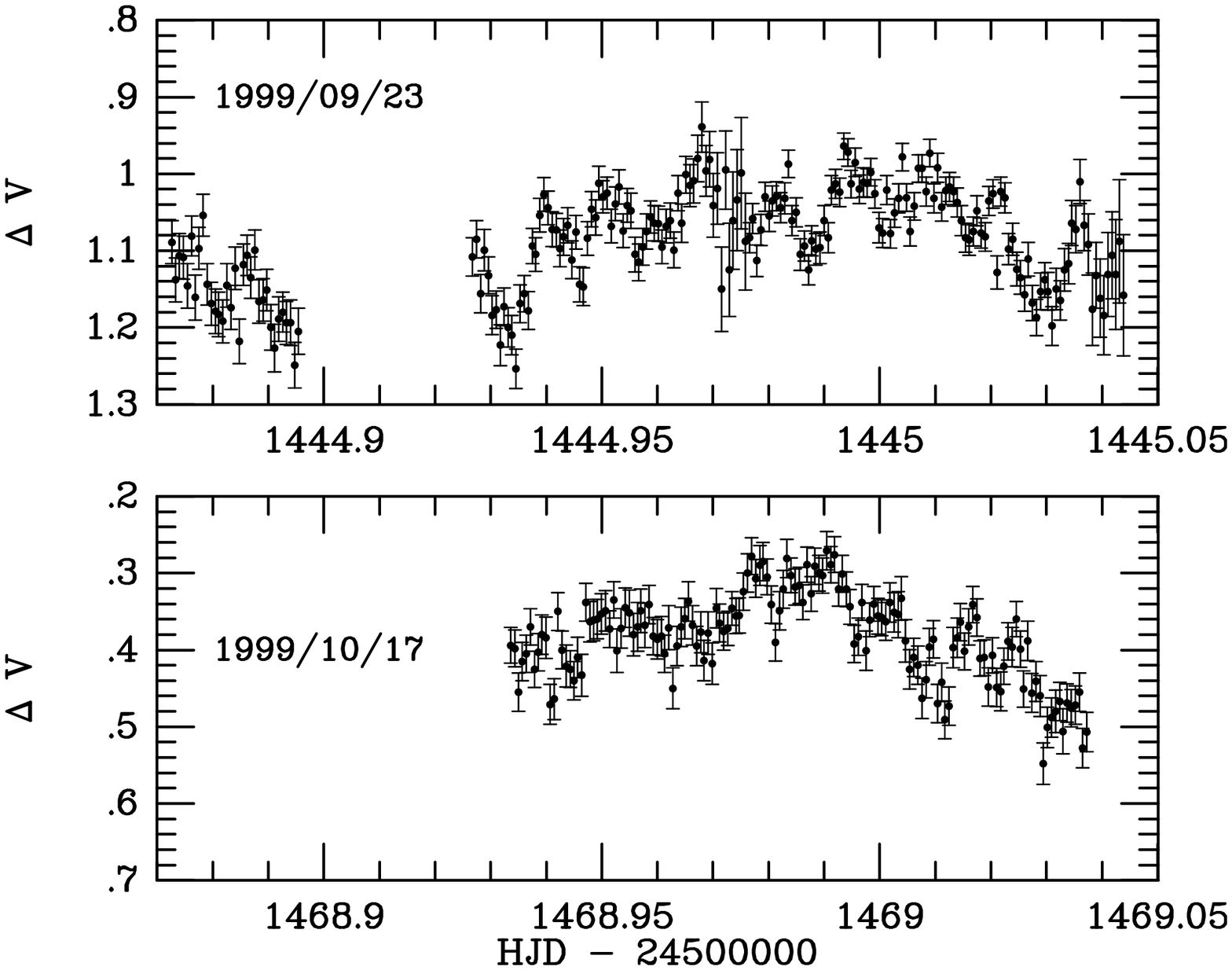,width=7in}
\end{figure}
\clearpage
\begin{figure}
\psfig{figure=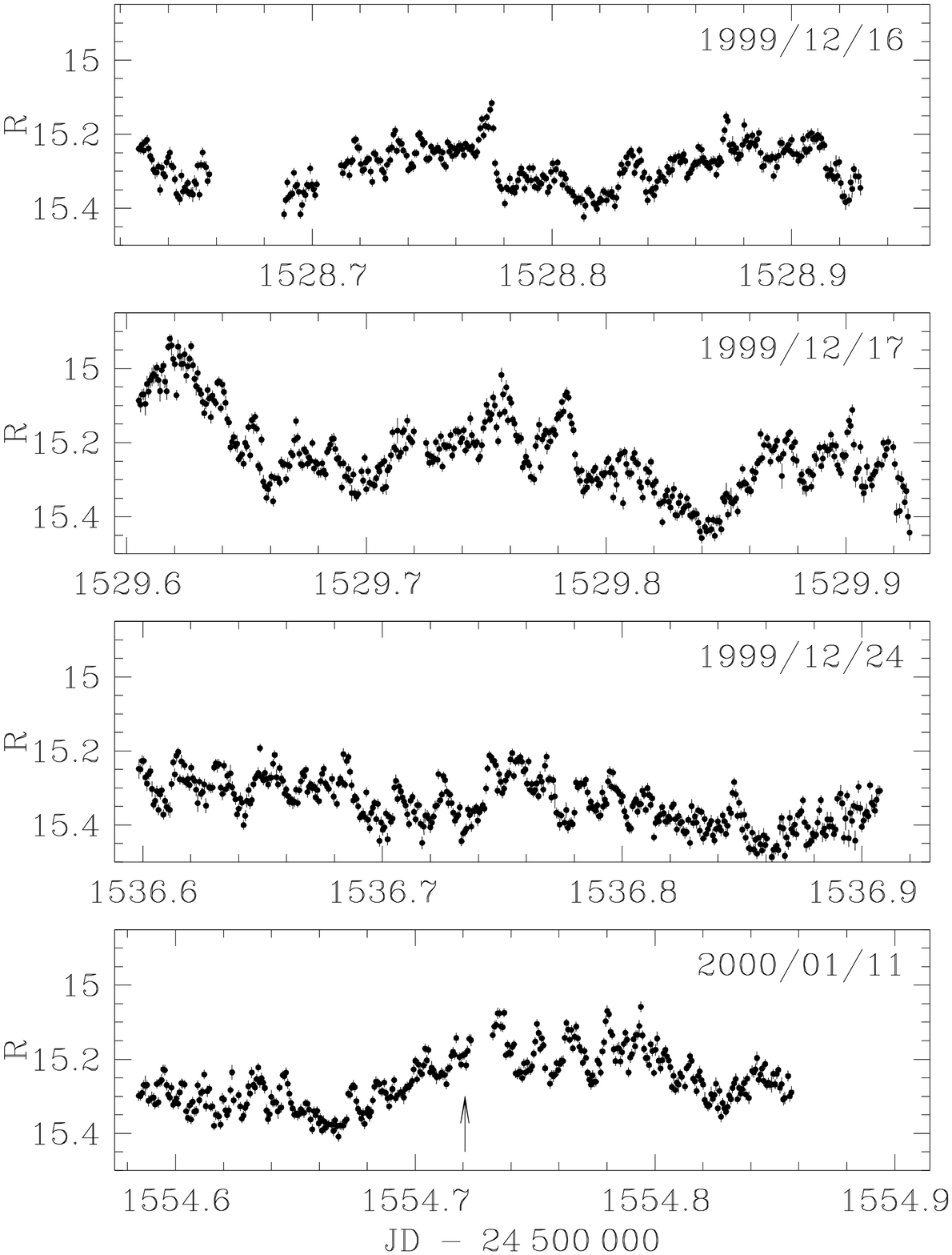,width=7in}
\end{figure}
\clearpage
\begin{figure}
\psfig{figure=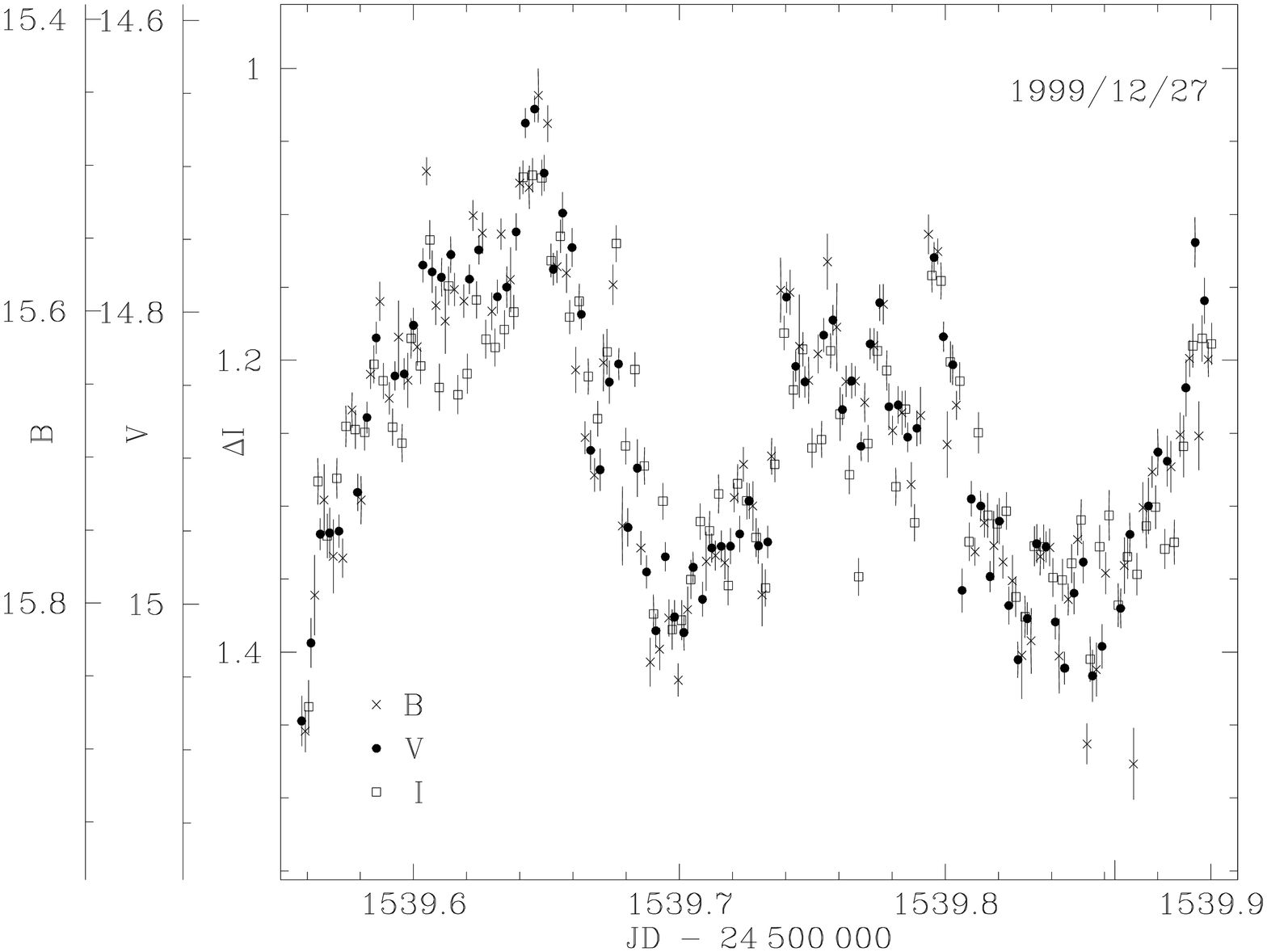,angle=270,width=7in}
\end{figure}
\clearpage
\begin{figure}
\psfig{figure=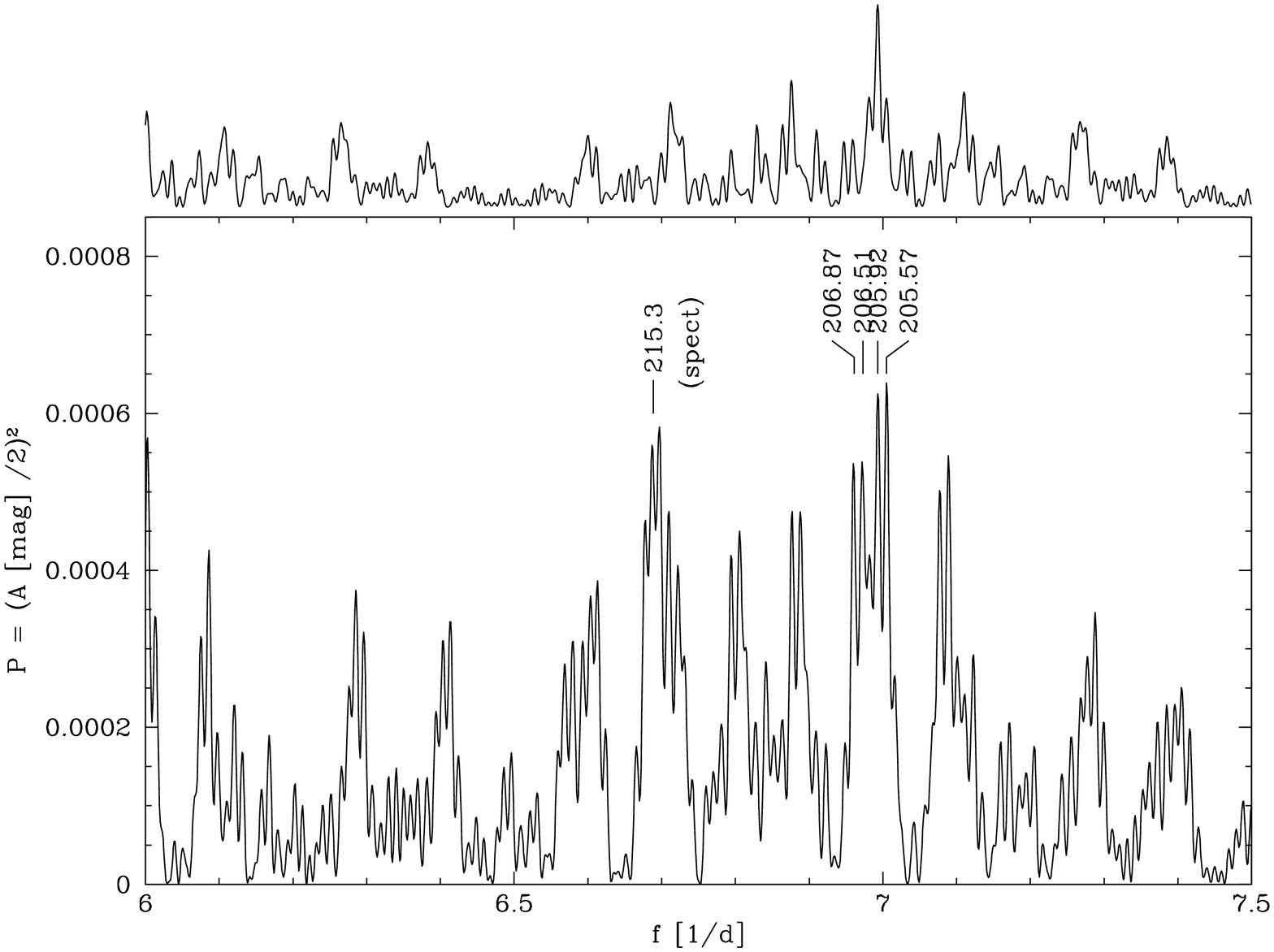,angle=270,width=7in}
\end{figure}
\clearpage
\begin{figure}
\psfig{figure=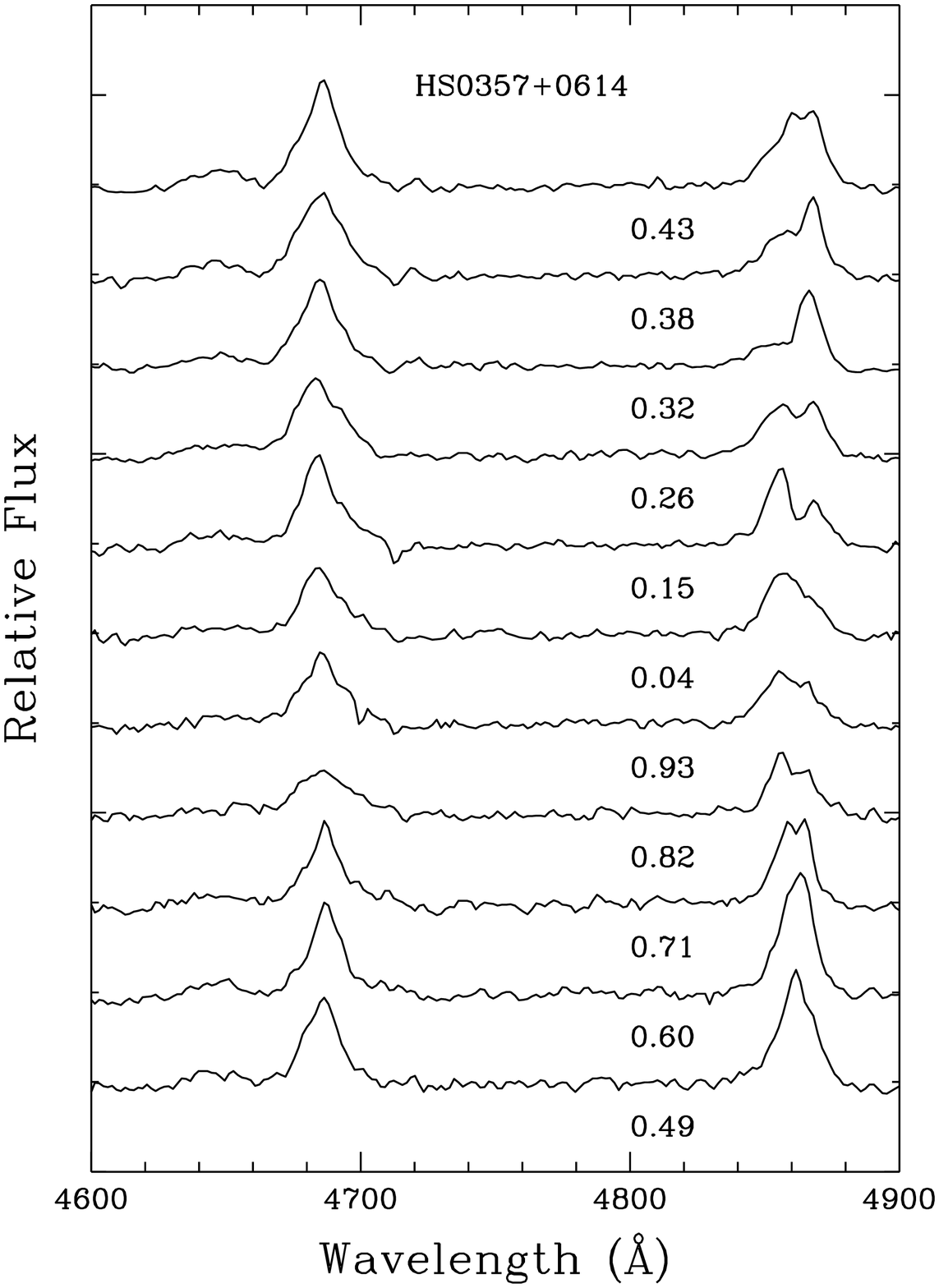,width=7in}
\end{figure}
\begin{figure}
\psfig{figure=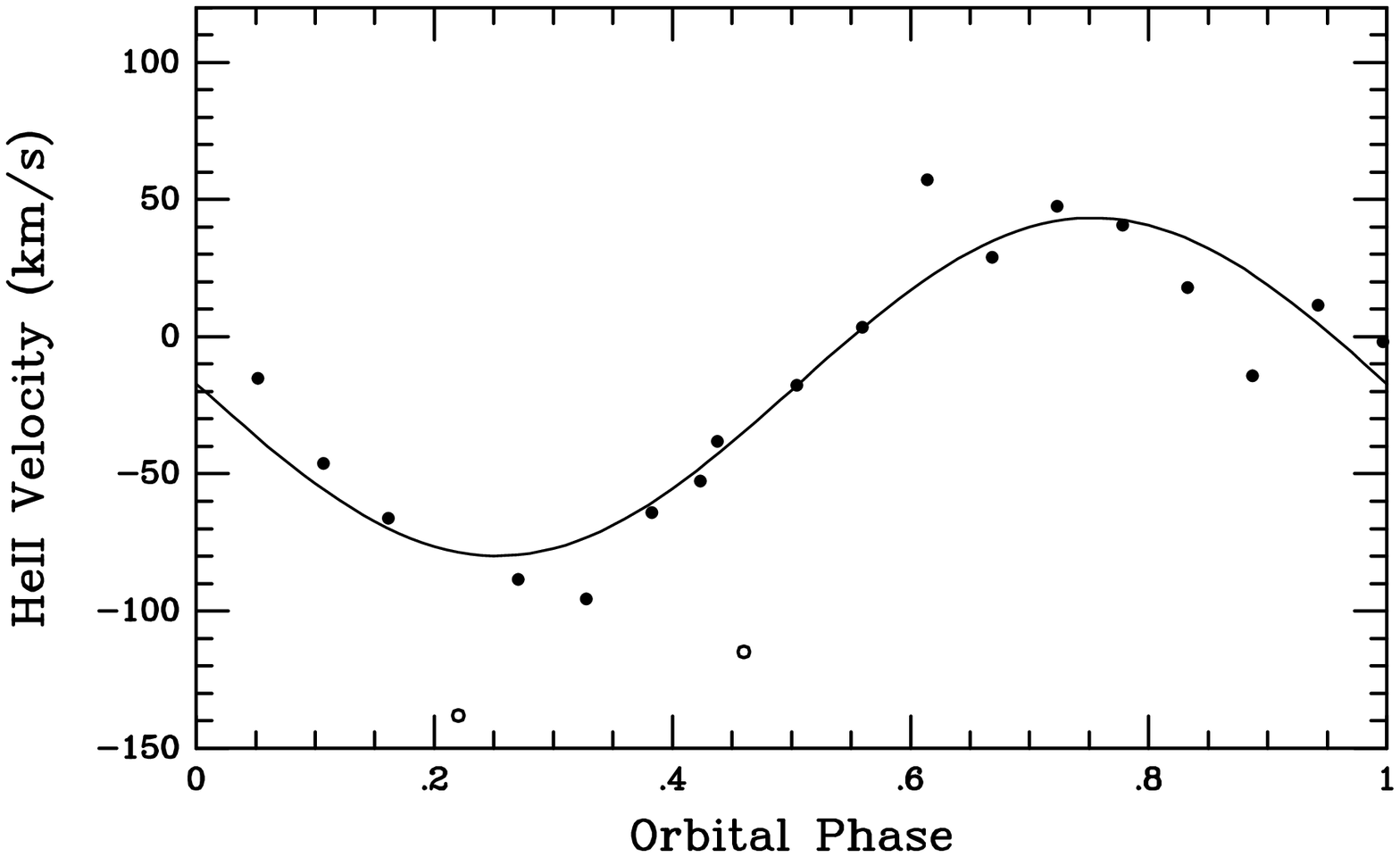,width=6in}
\end{figure}
\clearpage
\begin{figure}
\psfig{figure=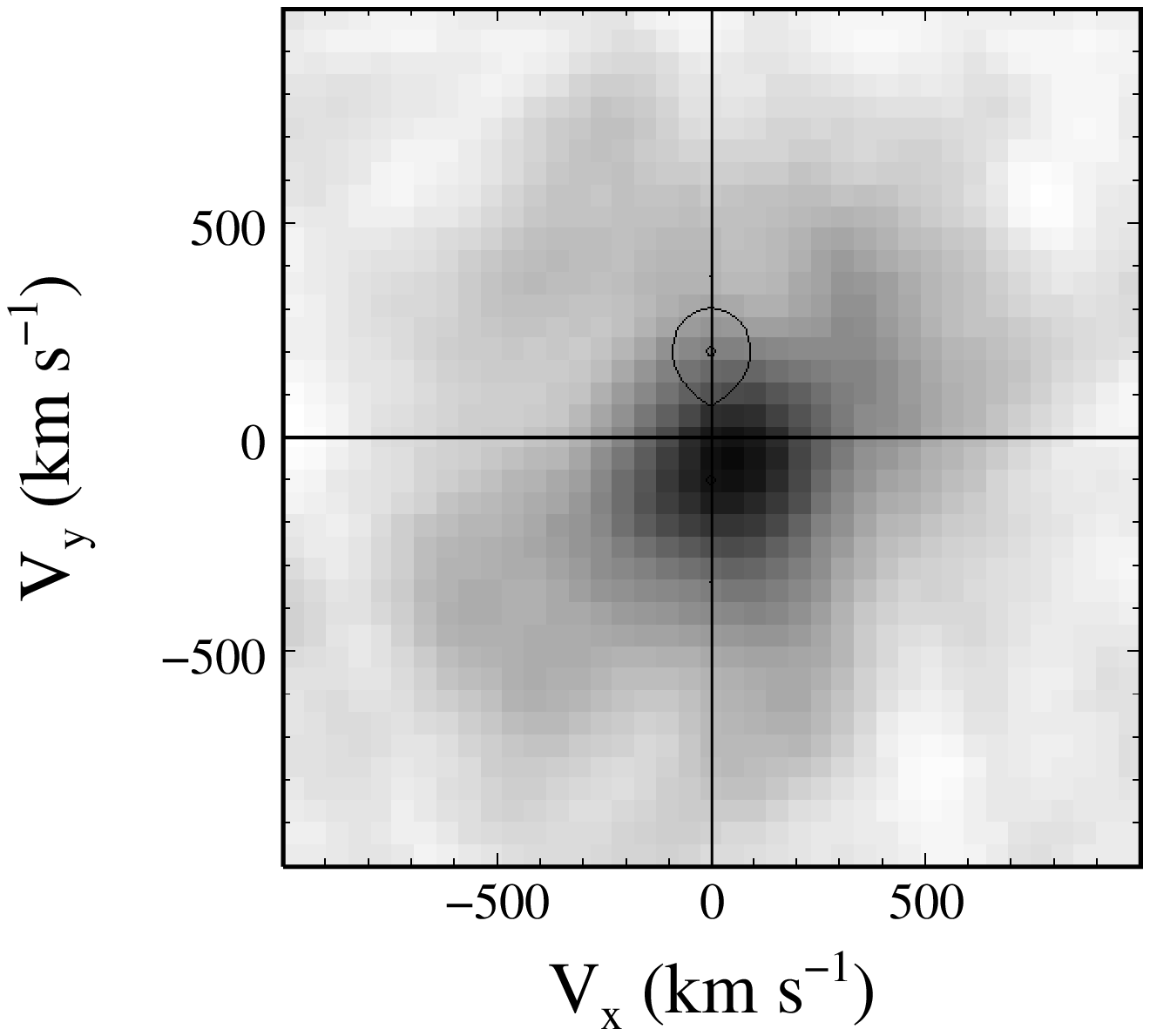,width=3.3in}
\psfig{figure=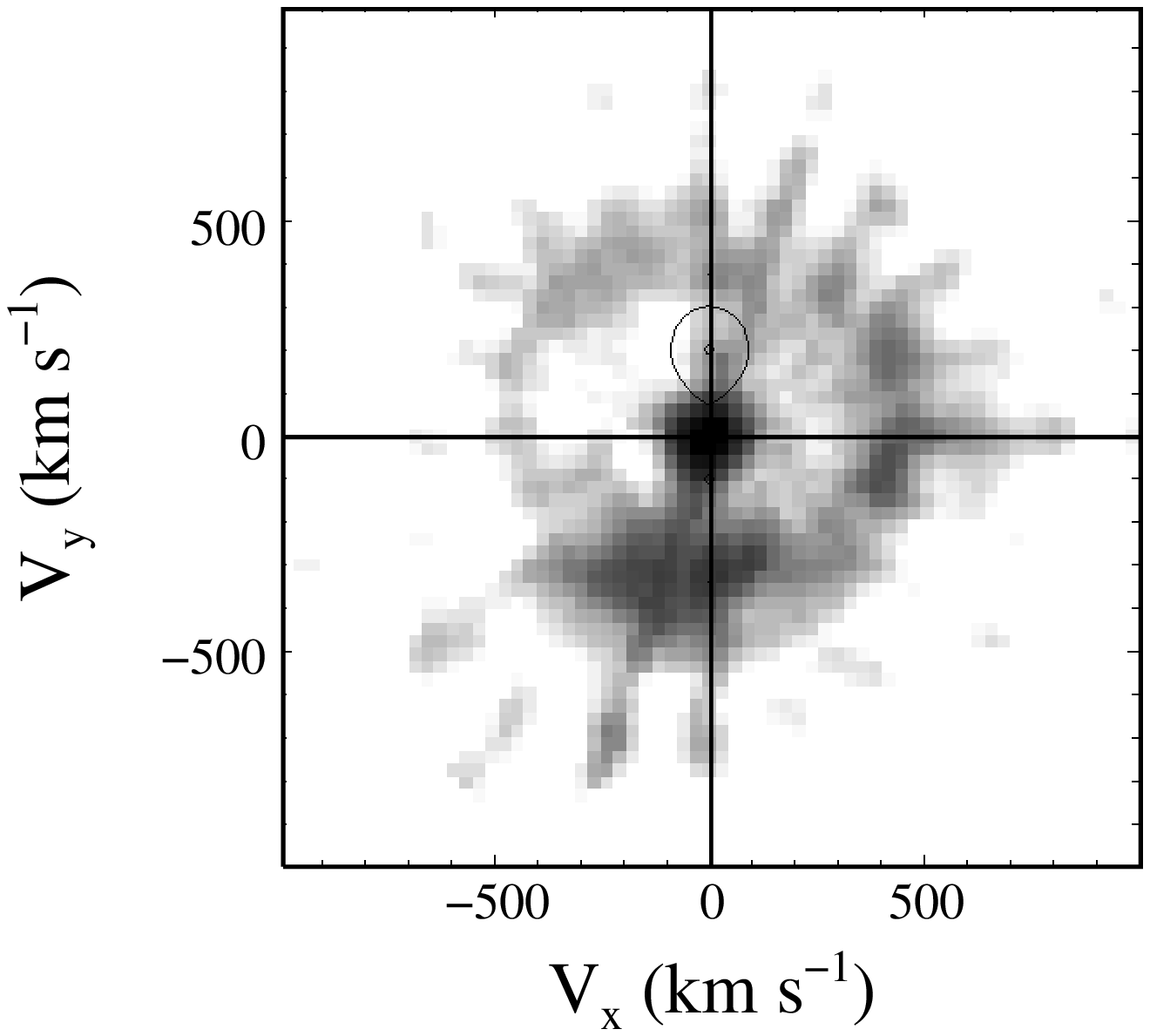,width=3.3in}
\psfig{figure=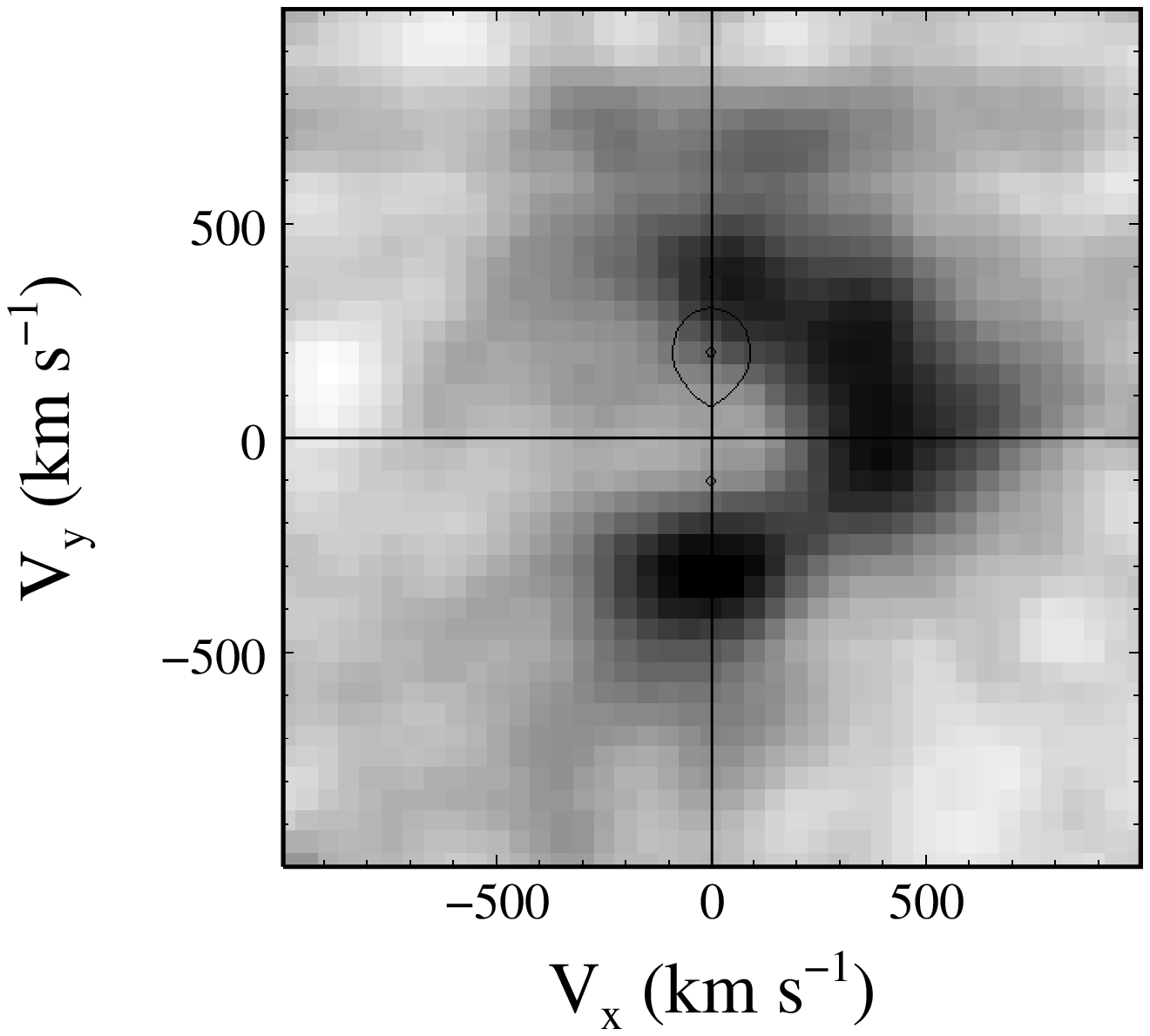,width=3.3in}
\end{figure}
\end{document}